\documentstyle[prl,aps,twocolumn,epsfig]{revtex}
\draft
\begin{document}

\twocolumn[
\hsize\textwidth\columnwidth\hsize\csname@twocolumnfalse\endcsname

\title{New results about the one dimensional Kondo lattice model}               
\author{Karyn Le Hur}
\address{Laboratoire de Physique des Solides, Universit{\'e} Paris--Sud,
		    B{\^a}t. 510, 91405 Orsay, France}
 \maketitle

 \begin{abstract}
 The one-dimensional Kondo lattice model (1D KLM) is usually defined by the 
 Kondo exchange $J$ between conduction electrons and spins of the array, and 
 the hopping strength $\text{t}$ for the moving electrons. Here, we also include a {\it direct} exchange $K$ term 
 between spins of the lattice, and we investigate situations where the bare value of $K$ does not exceed
 $J$. By using the non-Abelian bosonization, we show that a coherent heavy-fermion phase
 can be stabilized, when the bare value of $K$ exceeds the RKKY exchange (ruled by
 $T_A\simeq J^2/\text{t}$). Conversely, when $K<T_A$, the 
 low-energy fixed point is rather a Tomonaga-Luttinger (TL) liquid. 
 
  \end{abstract}

\vfill
 \twocolumn
\vskip.5pc ]
\narrowtext

  The Kondo lattice model (KLM) Hamiltonian 
  consists of conduction electrons antiferromagnetically coupled to
  a spin array {\it via} the Kondo interaction. This 
  model is usually defined by the two following parameters: the hopping strength $\text{t}$ of  
  conduction electrons and the antiferromagnetic Kondo coupling 
  $J$. Unlike the one impurity problem, poor man's scaling now generates 
  Ruderman-Kittel-Kasuya-Yosida (RKKY) interactions between the local
  moments. In actinide and rare-earth compounds, the bare 
  value of $J$ is very small: $J<<\text{t}$. It generally induces in rare-earth 
  compounds, an antiferromagnetic spin ordering transition at a temperature 
  $T_A\simeq J^2/\text{t}$. Indeed, the one-impurity Kondo energy scale 
  $T_k^{imp}\simeq\text{t}\text{e}^{\frac{-\pi t}{J}}$ cannot exceed the energy
  scale $T_A$, in that limit\cite{Do}; it would require a Kondo coupling constant, $\text{r}=J/t$ 
  of order unity. 
  \\
  However, certain rare-earth compounds such as $CeAl_3$, $CeCu_6$ or $CeCu_2Si_2$ do not become magnetic, at 
  low-energy; they rather yield a 
  heavy-fermion behavior\cite{po}. For $J<<\text{t}$, we may conclude that a heavy-fermion fixed
  point cannot occur in the KLM, {\it while} the RKKY process remains relevant. Thus, another magnetic parameter, more relevant 
  than the RKKY exchange, must be introduced in the bare theory to give a coherent heavy-fermion 
  behavior. To have a more precise idea about this parameter, we remember
  the well-known ``exhaustion'' phenomenon occurring in the 
  KLM\cite{no}. Even if the RKKY process is not relevant, only few
  conduction electrons may participate in the usual Kondo effect; we 
  just dispose of $n_c\simeq T_k^{imp}/\text{t}<<1$ electrons per site to screen the spins of the 
  array. In heavy-fermion systems, it implies that the singlet valence
  bonds between conduction electrons and spins of the array {\it cannot} be 
  prevalent. Then, since $J$ is supposed
  to be identical on each site, we realize that no renormalization process
  of $J$ can be relevant in that context. Finally, lattice spin effects should contribute to shift the Kondo energy scale $T_k$ to zero.
   \\
  Here, to describe heavy-fermion compounds, we add to
  the model, a nearest neighbour exchange $K$ ($K>0$) between the
  spins of the array; it is supposed to arise from {\it direct} exchange mechanisms between the spins of 
  the array\cite{po}. In rare-earth compounds, the bare value of $K$ should not exceed the
  local Kondo exchange $J$, but it is enough to distinguish two regimes. When the bare value 
  of $K$ tends to zero, the RKKY exchange gives indisputably a magnetic fixed point. But, when the {\it direct} exchange 
  mechanisms between local moments are the most prominent at energies 
  higher than $T_A$, the RKKY exchange does not play any role. Hence, {\it no} magnetic ground
  state can be stabilized in that case, due to the local Kondo exchange and 
  we suggest the occurrence of a heavy-fermion phase. Precisely, we
  predict, that the direct exchange K will renormalize to strong values at a ``high''
  temperature $T_{L}>>T_k^{imp}$. At lower energy, the spins of the lattice 
  are supposed to mutually screen one another; inevitably, the Kondo energy scale $T_k$
  tends to zero. For $T<<T_L$, the spin array resembles a resonating valence bond (RVB) system\cite{An}, and a coherent heavy-fermion system
  may occur at a low-energy scale $T_{coh}$. Indeed, the Kondo exchange 
  $J$ remains small, but it may introduce a small component of conduction electron single bonds to this RVB picture. It has been 
  claimed, that the wave-function describing such a spin system, shows evidence of a heavy-fermion fluid at the 
  Fermi-level\cite{PI}. Finally, we expect two different energy scales in the heavy-fermion phase of 
  the KLM.
  \\
  Theoretically, the recent development of field theory techniques applied to condensed matter have already
  allowed the investigation of the one-dimensional KLM (1D KLM)\cite{sept,huit,neuf,dix}. But, a heavy-fermion fixed point 
  and a magnetic one have not been found explicitly. In ref.\cite{quinze}, the
  Non-Abelian bosonization technique\cite{onze,douze,quatorze} has been used to show the emergence of a
  Haldane phase in the 1D KLM, for very large bare values of 
  $K$ ($K\simeq\text{t}$). Here, we show that it is still relevant to 
  discuss the occurrence of a coherent heavy-fermion fixed point and a 
  gapless phase too, in the extended 1D KLM (at half-filling):
 \begin{eqnarray}
 \label{zero}
 {\cal H}&=&-\text{t}\sum_{i,\sigma} c^{\dag}_{i,\sigma}c_{i+1,\sigma}+ (h.c) + U\sum_{i,\sigma}n^c_{i,\sigma}n^c_{i,-\sigma}\\ \nonumber
 &+&J\sum_i \vec{S}_{f,i}.c^{\dag}_{i}\frac{\vec{\sigma}}{2}c_i+ K\sum_i\vec{S}_{f,i}.\vec{S}_{f,i+1}\\ \nonumber
 \end{eqnarray}          
 For $T\simeq\text{t}$, we start with the bare parameters:  $J(t)<<\text{t}\ ,\ K(t)< J(t)\ ,\ U(t)>0$. Here, $c^{\dag}_{i,\sigma}$ $(c_{i,\sigma})$ creates (annihilates)
 an electron of spin $\sigma$ at site i and $\vec{S}_f$ is a spin $\frac{1}{2}$ 
 operator. A small U-Hubbard interaction between c-electrons ($U>0$) has 
 been introduced. The local exchange $J$ describes the usual Kondo effect. Here, we also consider a 
 nearest neighbour exchange K between the $S_f$ spins. The rare-earth compounds
 which are expected to cross towards a gapless ground state 
 obey, $K(t)\rightarrow 0$; those which are expected to cross towards a 
 heavy-fermion fixed point rather obey, $K(t)\simeq 2T_A$.
 
 Now, we use a continuum limit of the Hamiltonian (the lattice step $a\rightarrow 0$) 
 and we switch over to non-Abelian bosonization notations. We linearize the dispersion of conduction 
 electrons. On the Minkowskian space, the relativistic fermions
 $c_{\sigma}(x)$ are separated in left-movers $c_{L\sigma}(x)$ and
 right-movers $c_{R\sigma}(x)$ on the light-cone. We introduce the current operators
 for the charge and spin degrees of freedom, namely $J_{c,L}=:c^{\dag}_{L\sigma}c_{L\sigma}:$ and
 $\vec{J}_{c,L}=:c^{\dag}_{L\alpha}\frac{{\vec{\sigma}_{\alpha\beta}}}{2}c_{L\beta}:$ and similarly
 for the right-movers. The charge Hamiltonian is equivalently described in terms of the
 scalar field $\Phi_c^c$ and its moment conjugate $\Pi_c^c$\cite{5}:
  \begin{eqnarray}
 H_c&=&\int dx\ \frac{u_{\rho}}{2 K_{\rho}}:{(\partial_x\Phi_c^c)}^2:+\frac{u_{\rho}K_{\rho}}{2}: {(\Pi_c^c)}^2:\\ \nonumber
      &+&g_3\exp(i4k_F x)\cos(\sqrt{8\pi}\Phi_c^c)\\ \nonumber
 \end{eqnarray}
 The coupling $g_3\propto(Ua)$ generates the usual $4k_F$-Umklapp process; at half-filling, it makes the charge sector
 massive and the ground state insulating. The parameters $u_{\rho}$ and $K_{\rho}$, which 
 describe the Tomonaga-Luttinger (TL) liquid\cite{5}(or the Hubbard chain), are given by:
 \begin{equation}
 \label{n}
  u_{\rho}K_{\rho}=v \qquad \text{and} \qquad \frac {u_{\rho}}{K_{\rho}}=v+U/\pi
 \end{equation}
 The velocity $v$ is defined by $v=(ta)$. Concerning spin excitations in the 1D electron gas, the free boson Hamiltonian 
is written as:
 \begin{equation}
 \label{deux}
 H_s=\frac{2\pi v}{3}\int\ dx\ :\vec{J}_{c,L}(x)\vec{J}_{c,L}(x): + (L\rightarrow R)
 \end{equation} 
 For $K(t)<J(t)$, the free boson Hamiltonian controlled by $v_k=(Ka)$ is irrelevant. To bosonize the ``interchain'' spin 
 interaction, we need the complete bosonized representation for the conduction and the 
 localized spin operators, $\vec{S}_{f}$. They are given\cite{quatorze}:
  \begin{eqnarray}
  \label{s}
 c^{\dag}(x)\frac{\vec{\sigma}}{2}c(x)&\simeq&\vec{J}_{c,L}(x)+\vec{J}_{c,R}(x)\\ \nonumber
 &+&\exp(i2k_F x)\text{tr}(g.\vec{\sigma})\cos(\sqrt{2\pi}\Phi_c^c)\\ \nonumber
 \vec{S}_f&\simeq&\vec{J}_{f,L}(x)+\vec{J}_{f,R}(x)+(-1)^x\text{tr}(f.\vec{\sigma})
 \end{eqnarray}
 The spin operator $(-1)^x\text{tr}(f.\vec{\sigma})$ is not relevant 
 either because no {\it free} spinon can survive in the array due 
 the local exchange $J(t)>K(t)$. As expected, the single relevant
 interchain coupling is the Kondo term:
\begin{equation}
 H_k=\lambda_2\int dx\ \vec{J}_{c,L}(x)\vec{J}_{f,R}(x)+(L\leftrightarrow R)
 \end{equation}
 where:  $\lambda_{2}=(aJ)$. Finally, in the limit $K(t)<J$, the direct exchange K term between the 
 $S_{f}$ spins furnishes the following marginal interaction:
 \begin{equation}
 H_{coh}=v_k\int dx\ \vec{J}_{f,L}(x)\vec{J}_{f,R}(x)+(L\leftrightarrow R)
 \end{equation}
where $v_k=(aK)$. For $K(t)<J(t)$, it is the single authorized quadratic interaction between the 
$\vec{J}_{f}$ spinons. Such a procedure has been introduced in the zigzag spin chain problem\cite{neuf}. We
add, that $H_{coh}$ may be relevant {\it only} in the limit $K(t)<J(t)$; for $K(t)>>J(t)$, free excitations become
more prominent in the spin array\cite{neuf,dix,quinze}. 

{\it First}, we properly re-investigate the fixed point for $K(t)\rightarrow 0$. Here, $H_{coh}\rightarrow 
 0$ and the Kondo term $\lambda_2(t)>0$ is 
a marginal term. By using an Operator Product Expansion (OPE)\cite{douze,quatorze} of a SU(2) 
algebra, we confirm that $\lambda_2(t)$ might renormalize to strong values, producing the 
one-impurity Kondo energy scale:
 \begin{equation}
  T_k^{imp}[\lambda_2(t)]\propto \text{t}\text{e}^{-\frac{\pi v}{\lambda_2(t)}}
 \end{equation}
But in that case, the low-energy fixed point is not ruled by the energy scale $T_k^{imp}$. Indeed, in the limit $\lambda_2(t)<<v$, the RKKY process is obviously the
most relevant perturbative process: the RKKY energy scale $T_A\simeq J^2/t$ is much larger than $T_k^{imp}$. Hence, the low-energy fixed point 
is gapless and the term $\lambda_2\propto(Ja)$ is not renormalized at
low-energy: $\lambda_2(0)=\lambda_2(t)$. In 1D, we may not observe a ``true'' magnetic phase with a 
non-zero order parameter, but it is interesting to precisely discuss this 
gapless phase. Here, the fixed point can be subtly viewed as a TL 
liquid. Indeed, the spin Hamiltonian can be easily diagonalized, and we
obtain that spin excitations are still free: 
 \begin{equation} 
 H_s^*=\frac{2\pi v^*}{3}\int\ dx\ :\vec{J}_{L}'(x)\vec{J}_{L}'(x): + (L\leftrightarrow R)
 \end{equation}   
 But now, due to the term $\lambda_2$, the spinon field comes out as: 
  \begin{equation}
  \vec{J}_{L}'\simeq\cos\alpha\vec{J}_{c,L}+\sin\alpha\vec{J}_{f,R}
   \end{equation}
  While $\alpha\rightarrow 0$ or $\alpha\rightarrow\pi/2$, we can estimate that the ``composite'' spinon field $\vec{J}_{L}'$ still obeys a level-1 $SU(2)$ 
  algebra and, we have:
   \begin{equation}
   \tan\alpha=\frac{3\lambda_2(t)}{2\pi v}<<1
   \end{equation}
   It is due to the fact that only a small part of the 1D electron gas (of order $\sin\alpha$) is finally 
   trapped by the spin array; the chirality of c-electrons is not broken. This gapless phase resembles greatly a TL liquid, ruled inevitably by the charge field 
   $\Phi_c^c$ (since any charge fluctuation is allowed in the spin 
   array: $\langle\partial_x\Phi_f^c\rangle=0$) and by the spin field $\vec{J}'$. Hence, only the dynamical spin 
   properties of the 1D electron gas, are affected by the magnetic diffusion with the spin 
   array. The velocity of the free spinons is increased to $v^*=\frac{v}{\cos^2\alpha}\simeq v[1+\frac{9\lambda_2^2}{4\pi^2 v^2}]$, and the dynamical spin susceptibility 
   reads now\cite{5}:
   \begin{equation}
   \chi_s(\omega_n,q)=\frac{2}{\pi}\frac{q^2}{q^2v^*+\omega_n^2/v^*}
   \end{equation}
   Since $\alpha\rightarrow 0$, the f-electrons remain 
   localized and the Fermi-surface of the 1D electron gas is not really
   affected ($k_F=\pi/2a$). By hopping, the 1D electron gas just polarizes
   the spin array; the RKKY interaction may control the short-range distance
   behavior, making the characteristic structure around $q=2k_F$ in 
   correlation functions prominent. The definition of the ``composite'' spinon field 
   $\vec{J}'$ is very relevant to calculate the precise correlation functions in the
   spin array. Indeed, when $\lambda_2\not=0$, the localized-spin 
   correlation functions show a power-law with the same exponent as in the 1D electron gas\cite{5}:
   \begin{equation}
   \langle\vec{J}_{f,L(R)}(x)\vec{J}_{f,L(R)}(0)\rangle\simeq e^{i 2k_F x}\frac{\sin^2\alpha}{x^{1+K_{\rho}}}
   \end{equation}
   with, $\sin^2\alpha\simeq 9\lambda_2^2(t)/(4{\pi}^2 v^2)$. Finally, the stabilized TL 
   liquid phase allows to show evidence of the local-moment magnetism induced by 
   the large hopping term. This fact is independent of the band filling. 
    
{\it Second}, we investigate the fixed point, in the other case $K(t)\simeq 2T_A(t)$. Here, $H_{coh}$ becomes strongly relevant
and conversely, the RKKY exchange can be forgotten. It is now interesting to notice, that $H_{coh}$
has the same form as $H_k$. Both are marginal terms and then, renormalization equations of the
two parameters $v_k$ and $\lambda_2$ are finally strongly correlated. By using
the techniques of OPE, up to the second order, we obtain the two precise $\beta$-functions:
\begin{equation}
\frac{\partial\lambda_2}{\partial\text{ln}T}=\frac{-\lambda_2^2}{\pi(v+v_k)},\qquad
 \frac{\partial v_k}{\partial\text{ln}T}=\frac{-v_k^2}{\pi\lambda_2}
\end{equation}
For $\lambda_2(t)<<v$ and $v_k(t)\simeq 2aT_A(t)$, we may immediately observe 
that $\beta[v_k]>>\beta[\lambda_2]$. It shows, that the hopping term favors the renormalization of $v_k$ rather than the renormalization of $\lambda_2$, in the 
KLM. Using the $\beta[v_k]$-function, we obtain that $v_k$ renormalizes to strong couplings at the energy scale:
 \begin{equation}
 T_{L}[v_k(t),\lambda_2(t)]\propto\text{t}\text{e}^{\frac{-\pi\lambda_2(t)}{v_k(t)}}
 \end{equation}
 Integrating the first equation with bare parameters gives the 
Kondo energy scale, obtained in ref.\cite{huit}:
\begin{equation}
T_k^{\text{o}}[\lambda_2(t),v_k(t)]\propto\text{t}\text{e}^{\frac{-\pi[v+v_k(t)]}{\lambda_2(t)}}
 \end{equation}
Using Fig.1, we confirm that $T_L[2aT_A(t),\lambda_2(t)]>>T_{k}^{imp}$. From this point of view, $T_L$ 
is a {\it high} relevant energy scale, in the 1D KLM. Hence, for $T<<T_L$, we have to take into account  
 the strong renormalization process of $v_k(t)$; it implies that 
 $T_k^{\text{o}}[\lambda_2(t),v_k(t)]$ is not the precise expression for the
Kondo energy scale. The parameter $v_k$ obeys the thermal law:
\begin{equation}
v_k(T)=v_k(t)+\frac{v_k(t)^2}{\pi\lambda_2}\text{ln}\frac{T_{L}}{T}
\end{equation}
and the Kondo energy scale is rather
defined by the self-consistent equation: $T_k=T_k^{\text{o}}[\lambda_2(t),v_k(T_k)]$. In terms of bare 
parameters, it reads:
\begin{equation}
T_k^{\text{o}}[\lambda_2(t),v_k(t)]^{\frac{1}{1-\text{n}}}T_{L}[v_k(t),\lambda_2(t)]^{\frac{-\text{n}}{1-\text{n}}}
\end{equation} 
 with $\text{n}=\frac{v_k(t)^2}{\lambda_2(t)^2}$. As shown in Fig.1, it is reduced
 to zero when $v_k(t)\rightarrow 2aT_A$. Remarkably, the large renormalization of $v_k$ 
 implies that $\lambda_2(t)$ remains a constant parameter in the 
 model. Singlet valence bonds formed between {\it two} $S_f$ spins will be the most 
 prominent in the KLM and the ``exhaustion'' phenomenon does not occur as a real 
 problem.

  \begin{figure}
    \centerline{\epsfig{file=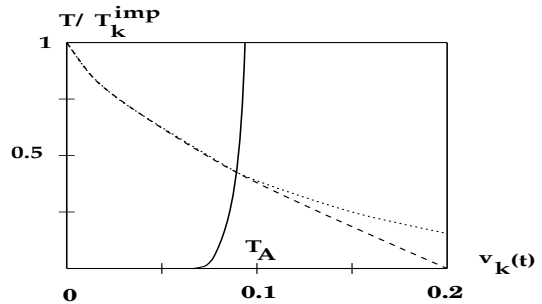,angle=0.0,height=4cm,width=7cm}}
    \vskip 0.2cm
    \caption{ Plots of the lattice energy scale $T_{L}$ (full line), the
     Kondo energy scale $T_k^{\text{o}}$ (dotted line), and
     the Kondo energy scale $T_k$ (dashed line), versus the bare coupling 
    $v_k(t)$; other parameters are given:\ $\lambda_2(t)=0.3$, $a=1$, $t=1$, $T_A=0.1$.}
    \end{figure} 
 
 $T_{L}$ corresponds to the onset of strong very-short range spin 
 correlations in the array with singlets bonds formed between the $S_f$ spins. Now, we begin to investigate the low-energy behavior (for $T<<T_L$), in the case where $J=0$. To
 simplify the calculations, we consider that $v_k(T)$ behaves as a large constant parameter
 $v_k^{\text{o}}$ of order $v$ when $T<<T_L$. For $J=0$, this massive phase is characterized by an order parameter $d=\langle\vec{S}_{f,i-1}.\vec{S}_{f,i}\rangle-\langle\vec{S}_{f,i}.
 \vec{S}_{f,i+1}\rangle$, which varies linearly with $T_{L}$. Unlike the gapless phase, we just expect a short-range N\' eel order, with a correlation 
 length $\xi_{L}\propto T_{L}^{-1}$:
   \begin{equation}
   \langle\vec{S}_{f,i}.\vec{S}_{f,j}\rangle\propto (-1)^{i-j}\text{e}^{-\left| i-j\right|/\xi_{L}}
   \end{equation}
  In the array, singlet bonds can be formed between two $S_f$ spins over a distance of 
  $x\leq\xi_L$. Since $\xi_L>>a$, the spin array resembles a RVB state, formed by the 
  $S_f$ spins only. The RVB wave-function is known to be very similar to 
  the well-known BCS wave-function\cite{An}. From this point of view, this state is seen to be the projection of 
  a free Fermi sea of f-electrons, and the theory reflects now the existence of two {\it independent}
  Fermi surfaces (FS); both are characterized by $k_F=\pi/2a$. Here, the f-electrons liberate one another {\it close} to the
  Fermi-energy ($E_f\rightarrow 0$).
 \\
 Now, we must discuss the influence of a small $J$ coupling on these 
 two Fermi surfaces; it is convenient to use the spin 
 formalism. In the limit $\lambda_2<<v$, many free spin excitations will 
 survive in the 1D electron gas at low-energy, and the 
 Hamiltonian (\ref{deux}) remains relevant at the fixed 
 point. But here, the f-electronic fluid resonates near the Fermi-level, and finally
 the small $J$ local coupling may affect the above RVB 
 description. The low-energy behavior is also described by the following Hamiltonian: 
 \begin{eqnarray}
\label{mu}
 H_{coh}^*&=&v_k^*\int\ dx\ \vec{J}_{L}''(x)\vec{J}_{R}''(x)+(L\leftrightarrow R)\\
   \label{jo}
   \vec{J}_{L}''&\simeq&\sin\beta\vec{J}_{f,L}+\cos\beta\vec{J}_{c,L}
   \end{eqnarray}
   Unlike the gapless phase, the angle $\beta$ tends to $\pi/2$:
   \begin{equation}
   \tan^{-1}\beta=\frac{\lambda_2}{v_k^{\text{o}}} <<1 
  \end{equation}
  By using Eqs(\ref{mu}),(\ref{jo}), we conclude that $H_{coh}^*$
  describes a sea of singlet resonating valence bonds formed by the 
  $S_f$ spins, with a small admixture of conduction electron single 
  bonds. The coupling $v_k^*=v_k^{\text{o}}[1+\frac{\lambda_2^2}{v_k^{\text{o}2}}]$ is still 
  larger than $v_k^{\text{o}}$ so this RVB state is very stable. 
  \\
  Now, it is relevant to investigate the
  corresponding wave-function of such a RVB state. We take a Abrikosov (fermionic)
  representation of the ``composite'' spinon $\vec{J}''$. Then, we observe that the small local exchange $\lambda_2$ hybridizes
  each f-state with a small component of conduction electron. The RVB wave-function 
  reads now:
  \begin{equation}
  |RVB\rangle=\prod_{\vec{q}}\ \psi_{\vec{q}
  \uparrow}^{\dag}\psi_{-\vec{q}\downarrow}^{\dag}\ |O\rangle
  \end{equation}
  When $\beta\rightarrow\frac{\pi}{2}$, we have: $\psi_{\vec{q}
  \sigma}^{\dag}|O\rangle\simeq[\cos\beta c^{\dag}_{\vec{q}\sigma}+
  \sin\beta f^{\dag}_{\vec{q}\sigma}]|O\rangle$; $|O\rangle$ is the vacuum. For
  a finite lattice step, we have used the isomorphism: $\vec{S}'_{\vec{q}}\vec{S}'_{-\vec{q}}\leftrightarrow \psi_{\vec{q}
  \uparrow}^{\dag}\psi_{-\vec{q}\downarrow}^{\dag}$. First, the $\psi$-energy
  level can be identified with the f-energy level: $E_{\psi}\simeq E_f/\sin^2\beta
  \simeq E_f$. But now, we can check that the f-level is enlarged to the characteristic energy scale:
  \begin{equation}
  T_{coh}\propto\frac{E_f^2 V^2}{\text{t}}\hskip 0.4cm\text{with}\hskip 0.4 cm V=\tan^{-1}\beta
  \end{equation}
  This expression of $T_{coh}$ resembles the expression found by
  mean-field studies of the well-known Anderson lattice model\cite{Piers}. $T_{coh}$ defines the 
  onset of heavy-fermion behavior in the
  1D KLM. It comes from the enlargement of the (Fermi) f-level by the Kondo exchange:
  \begin{equation}
    F(T)=F_o(T)+\frac{N_f}{\pi}\int\ d\omega\ f(\omega)[\tan^{-1}(\frac{T_{coh}}{\omega})]
    \end{equation}
    $F_o(T)$ is the free energy of the 1D electron gas, $f(\omega)$ is the Fermi distribution function
    and $N_f$ is the number of f-electrons. The TL liquid behavior characterized
  by the specific heat, $\frac{C}{T}\propto\frac{\partial^2 F_o}{\partial T^2}
  \simeq\frac{v}{u_\rho}$ becomes less prominent than the Fermi liquid behavior characterized by the huge $\frac{C}{T}\propto
  \frac{N_f}{T_{coh}}$. Second, at half-filling, the number $N_{\psi}$ of free ``composite'' fermions 
  satisfies the precise Luttinger's sum rule: 
  \begin{equation}
  k_F=\frac{N_{\Psi}\pi}{2}=\frac{\pi}{2}(N_c\cos^2\beta+N_f\sin^2\beta)
  \end{equation}
  The length of the lattice is fixed to $L=1$. It yields the result that the Fermi-surface of the composite fermions also obeys $k_F=\pi/2a$. Third, since the 
  Kondo exchange $J$ remains small, {\it no} phase shift occurs in the wave-function of the c-electrons: the Friedel 's sum rule 
  is reduced to the trivial law, $\delta=0$. Finally, it is enticing to remark the narrow link between the RVB
  insulating state description and the occurrence of a narrow band (of width $T_{coh}$ and length $2k_F=\pi/a$) at the Fermi level.
   
  Summarizing, we have shown the occurrence of a
  TL liquid phase as well as a heavy-fermion fixed point in the 1D KLM, for
  small bare values of $K$ ($K(t)<<J(t)$). The cross-over, between these
  two new Kondo insulators, occurs when $K(t)$ has the same strength as
  the RKKY exchange. Now, it would be interesting to give a complete $K$-phase diagram of the 1D KLM.

 \end{document}